\documentclass[%
 reprint,
superscriptaddress,
showpacs,preprintnumbers,
 amsmath,amssymb,
 aps,
prl
]{revtex4-1}
\usepackage{dcolumn}
\usepackage{float}
\usepackage{graphicx}
\usepackage{dcolumn}
\usepackage{bm}
\usepackage{epstopdf}
\usepackage{mathtools}
\usepackage{physics}

\usepackage{hyperref}

\begin{document}

\title{Spin selective scattering modes in random anisotropy optical medium
}
%
\author{Ankit Kumar Singh}\email{aks13ip027@iiserkol.ac.in}
\affiliation{Indian Institute of Science Education and Research Kolkata, Mohanpur 741246, India}

\author{Antariksha Das}
\affiliation{Indian Institute of Science Education and Research Kolkata, Mohanpur 741246, India}
\affiliation{QuTech, Delft University of Technology, 2628 CJ Delft, The Netherlands}
\author{Sourin Das}
\affiliation{Indian Institute of Science Education and Research Kolkata, Mohanpur 741246, India}
\author{Nirmalya Ghosh}
\affiliation{Indian Institute of Science Education and Research Kolkata, Mohanpur 741246, India}
%


\begin{abstract}
Geometrical and dynamical phase have competing effects as far a scattering of light form inhomogeneous anisotropic optical medium is concerned. If fine-tuned appropriately, these effects can completely cancel each other for a chosen spin component while having an additive effect on the orthogonal components. Here, we show a manifestation of extraordinary spin selective modes in Fourier spectrum of Gaussian beam transmitted through anisotropic disordered medium. We realize the concept using a twisted nematic liquid crystal-based spatial light modulator (SLM) with random gray level distribution for incident Gaussian beam.


\end{abstract}

\maketitle

Coupling of intrinsic polarization of light with its orbital degrees of freedom due to external medium goes by the name of spin-orbit interaction (SOI) of light and is a topic of recent interest \cite{bliokh2015spin,ma2016spin,shitrit2013rashba,shitrit2013spin,gerd_2014,lin2014dielectric,ling2014realization,ling2015giant,pal2016tunable,ling2017recent,hosten2008observation,maguid2017disorder,frischwasser2011rashba}. An effective Hamiltonian that dictates the evolution of the spin degree of freedom of light in an anisotropic and inhomogeneous dielectric medium with can be expressed as $\hat{\Delta}/2-\lambda\hat{A}.\mathbf{\dot{p}}/2\pi$, where $\mathbf{\dot{p}}$ is the derivative of momentum operator $\mathbf{p}$, $\hat{\Delta}$ denotes the anisotropicity of the medium, $\hat{A}$ is the Gauge potential, and $\lambda$  is the wavelength of light \cite{PhysRevA.75.053821,ma2016spin,shitrit2013spin,frischwasser2011rashba,wilczek1984appearance,zak1989berry,de1998noncyclic}. The first term in the Hamiltonian can be identified as the Zeeman term while the second term is associated with the SOI of light. Some of the very interesting effects arising of such Hamiltonians which may be analogous for electrons in condensed matter systems are spin Hall effect of light from spatially tailored medium, the optical Rashba effect, and the spin-dependent beam shift, etc. \cite{bliokh2015spin,ma2016spin,shitrit2013rashba,shitrit2013spin,gerd_2014,lin2014dielectric,ling2014realization,ling2015giant,pal2016tunable,ling2017recent,hosten2008observation}. It should be noted that the interpretation of the Zeeman term and the SOI term in the Hamiltonian can be related to the dynamical and geometrical phase of light respectively, for a cyclic as well as a non-cyclic process \cite{wilczek1984appearance,zak1989berry,de1998noncyclic}.

Most of the cases discussed above are of synchronous ordered system. In this regard, it was recently shown that for a completely disordered inhomogeneous anisotropic optical system, one can get a very interesting manifestation of such spin optical effect with random spin-split scattering modes over the entire momentum domain, which can be characterized as the random optical Rashba effect \cite{maguid2017disorder}. The origin of such an effect can be interpreted as the disorder in the second term of the Hamiltonian or equivalently the geometric phase distribution. 

Here, we report a remarkable effect that can originate from a perfect synchrony between the Zeenam type and term and the SOI term though individually they are fully disordered and random. Such a disordered anisotropic systems have the potential to demonstrate input spin selective/asymmetric random scattering modes i.e., the randomly scattered modes are observed for one spin states while the other state follows an entirely different trajectory with almost no effect of the disorder in the system. The spin controlled active tuning of the random scattering modes in momentum space may provide a route for spin-controlled applications in spintronics and nanophotonics based on such disordered anisotropic medium \cite{manchon2015new,jendrzejewski2012three,volpe2014speckle,chaigne2016super,gateau2013improving}. We have demonstrated this idea using an optical system, however, the finding are general and may have useful applications in various optical, quantum and condensed matter systems.
When a circularly polarized Gaussian beam ($G(x,y)=e^{-(x^2+y^2)/w_o^2}$, $w_o$ is the beam width) propagates through a spatially inhomogeneous anisotropic medium, it acquires a phase distribution ($\phi(x,y)$), creating a spatial phase gradient in transverse to the beam propagation direction \cite{bliokh2015spin,shitrit2013spin,ling2017recent,lin2014dielectric,ling2015giant,ling2014realization,maguid2017disorder,pal2016tunable}. The electric field ($\mathbf{E_t}$) transmitted from the inhomogeneous medium can be written as
\begin{equation}
\mathbf{E_t}=e^{i\phi(x,y)}G(x,y)\ket{+/-}
\end{equation}
Here, $\ket{+/-}$ indicate the incident right/left circular polarized (RCP/LCP) polarization state of light and the phase $\phi$ ($=\pm\phi_g(x,y)+\phi_d(x,y)$) is the spatially varying total phase [sum of the dynamical ($\phi_d$) and the geometrical phases ($\pm\phi_g$, $+(-)$ for RCP (LCP) state of light)] acquired by the circularly polarized field. The violation of spatial inversion symmetry of the system by the inhomogeneous distribution of the total phase leads to the SOI of light \cite{bliokh2015spin,shitrit2013spin,ling2017recent,lin2014dielectric,ling2015giant,ling2014realization,maguid2017disorder,pal2016tunable}. The strength of such an effect depends on the phase inhomogeneity of the light beam, for a simple case of where $\phi=cx$ (for some constant $c$), a shift proportional to “$c$” is observed between the two circular polarization state in the momentum space. In case the total phase ($\phi=\pm\phi_g$) is entirely geometrical and varies randomly in space \cite{maguid2017disorder}, a complex transverse momentum ($\mathbf{k_\perp}$, $k_x$ and $k_y$) distribution of the field can be observed with the $\mathbf{k_\perp}$ values distributed throughout the momentum space as
\begin{equation}
	\label{eq2}
	I_t(k_x,k_y)=\left|\iint \limits_{-\infty}^{+\infty} e^{-i(k_x x+k_y y)}\mathbf{E_t}(x,y)dx dy\right|^2
\end{equation}
The above equation corresponds to the intensity of spin-orbit coupled \textit{symmetric} random spin-split scattering modes (i.e. $I_t^{LCP}(k)=I_t^{RCP}(-k)$) in the momentum space due to opposite sign of the geometrical phase ($\phi_g(x,y)$) acquired by the LCP and RCP polarizations \cite{bliokh2015spin,shitrit2013spin,ling2017recent,lin2014dielectric,ling2015giant,ling2014realization,maguid2017disorder,pal2016tunable}. However, the dynamical phase does not show such dependence on the polarization of light \cite{ngbook}. Thus a spatially varying geometrical and dynamical phase would give a spin-orbit coupled \textit{asymmetric} randomly scattered spin-split modes (i.e. $I_t^{LCP}(k)\neq I_t^{RCP}(-k)$). The spin asymmetricity ($I_{sa}$) random scattering modes in such system can be quantified as
\begin{equation}
	I_{sa}=\left<\dfrac{\left|I_t^{LCP}(k_x,k_y)- I_t^{RCP}(-k_x,-k_y)\right|}{I_t^{LCP}(k_x,k_y)+ I_t^{RCP}(-k_x,-k_y)}\right>
\end{equation}
$\left<...\right>$ denotes a sum over all the $k_\perp$ values.

The SOIs from such disordered inhomogeneous anisotropic medium can also be understood in terms of the spatial autocorrelation function ($A(x,y)$) of the electric field transmitted from the disordered medium that can be related to the momentum distribution of the field intensity (power spectrum) through \textit{Wiener-Khinchin theorem} as
\begin{equation}
\label{eq4}
I_t(k_x,k_y)=\iint \limits_{-\infty}^{+\infty} e^{-i(k_x x+k_y y)}A(x,y)dx dy
\end{equation}
It implies that, a complex-valued autocorrelation function that changes randomly in space has all the spatial frequency components with varying amplitude, leading to the random scattering modes in the Fourier plane.
\begin{figure}[ht]
	\centering
	\includegraphics[width=0.9\linewidth,keepaspectratio]{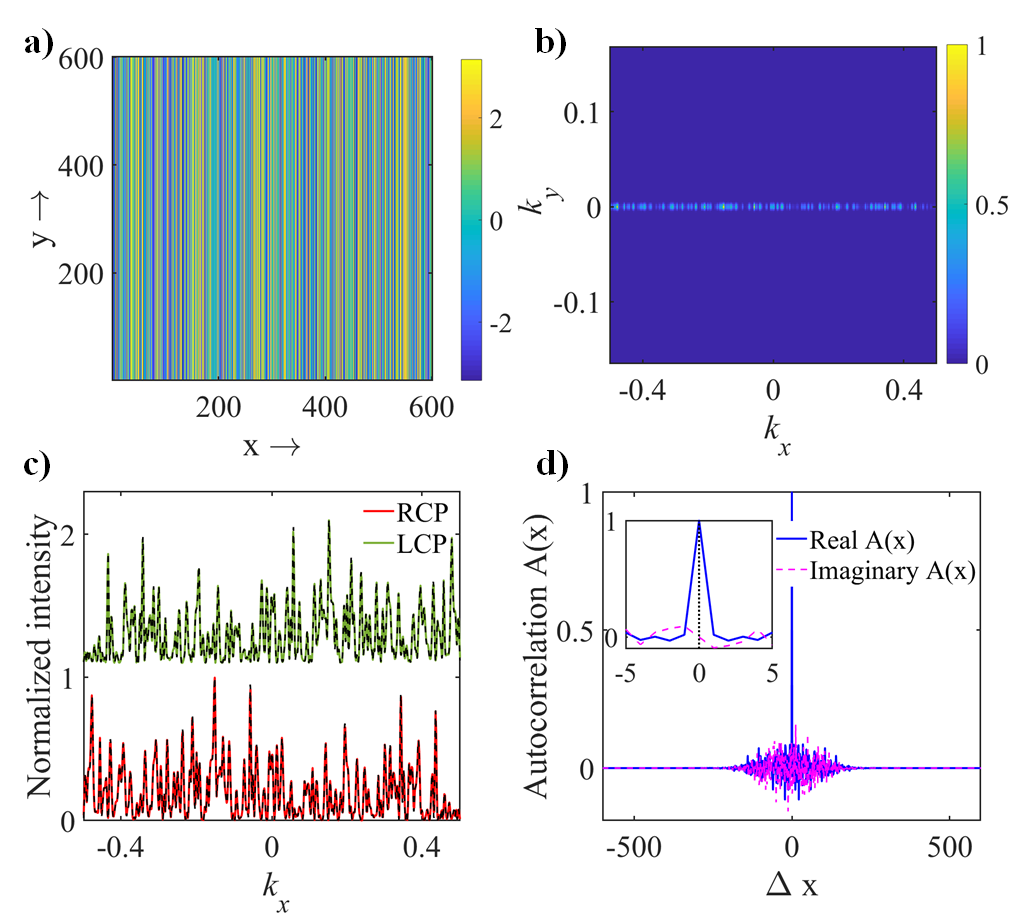}
	\caption{\label{fig1}(a) The geometric phase distribution of the anisotropic medium  disordered along the $x$-direction and, (b) the momentum space (normalized with wavevector ($k_o$) of the incident beam) intensity distribution showing the random scattering modes for (b) RCP polarization state, (c) LCP and RCP polarization states along $k_x$ with $k_y=0$ showing the spin symmetric random scattering modes [calculated using Eq.\ref{eq2} (solid line) and Eq.\ref{eq4} (dashed line)], the intensity of LCP state is shifted for better visualization. (d) The symmetric real and the anti-symmetric imaginary parts of the autocorrelation function of the Gaussian beam transmitted from the disordered anisotropic medium, the  inset shows a zoomed plot around $\Delta x=0$.}
\end{figure}
\begin{figure}[ht]
	\centering
	\includegraphics[width=0.65\linewidth]{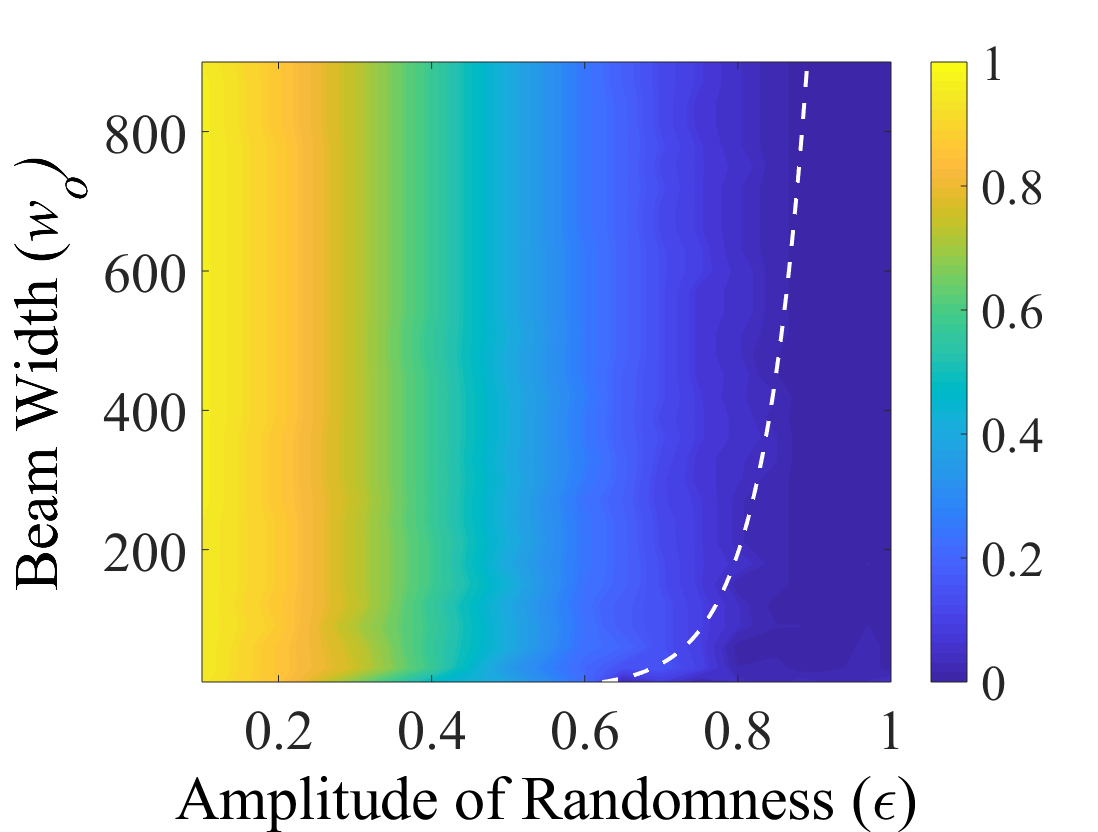}
	\caption{\label{fig2}The effect of the beam width of the input Gaussian beam $w_o$ and the amplitude of randomness ($\epsilon$) in the disordered anisotropic system on the difference between the maximum of the real and imaginary parts of autocorrelation function is shown in the color bar. The dotted line marks transition from usual momentum domain spin-Hall effect (above the line) to random spin split modes (below the line).}
\end{figure}
\begin{figure*}[ht]
	\centering
	\includegraphics[width=0.7\linewidth]{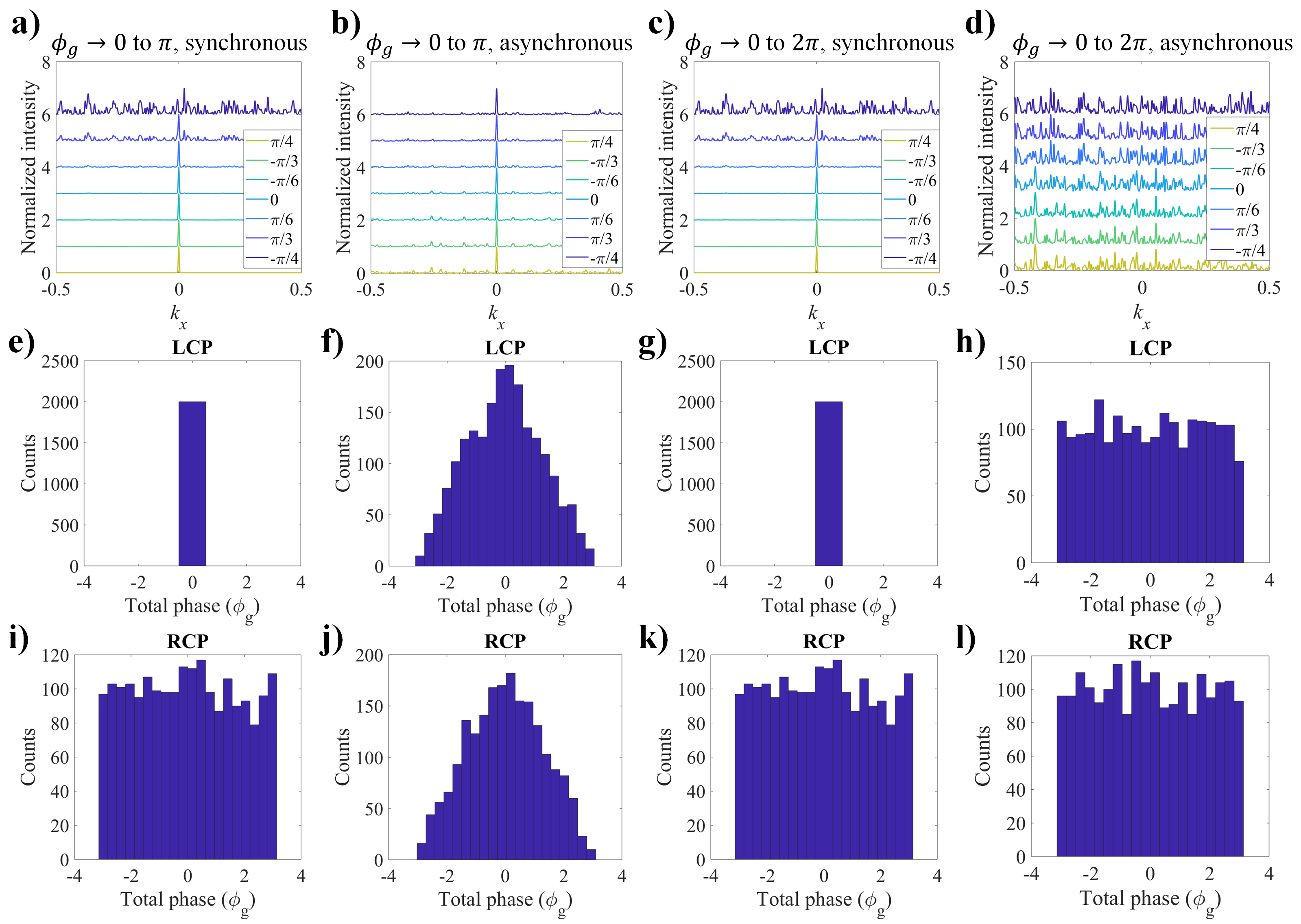}
	\caption{\label{fig3}(a-d) The elliptical polarization states ($[1,e^{i\delta} ]^T$, ellipticity ($\delta/2$) is mentioned in legends) were used to actively tune the momentum space intensity distribution from the random scattering modes to a Gaussian mode (the intensities are shifted for better visualization). The spatially random system was taken such the geometric phase ($\pm\phi_g (x)$,$+$($-$) for RCP (LCP) polarization of light) of the transmitted light is uniformly distributed between (a,b) $0$ to $\pi$ and (c,d) $0$ to $2\pi$. The dynamical phase acquired in the evolution is (a,c) synchronous ($\phi_d(x) = \phi_g (x)$), (b,d) asynchronous ($\phi_d(x)$ is distributed in same manner as $\phi_g(x)$), to the geometrical phase. (e-l) The total phase frequency distribution for LCP and RCP states are given below the  corresponding normalized intensity distributions of the disordered medium.} 
\end{figure*}

In Fig.\ref{fig1}, we have shown a two-dimensional spatial distribution of geometric phase ($\phi_g$) which is obtained by a RCP Gaussian beam ($w_o=80$ arb. units), when passed through the spatially disordered anisotropic medium with randomized phase gradients. The geometric phase ($\phi_g(x)$) is chosen to be a uniformly distributed random function of x coordinate [$-\epsilon\pi\leq \phi_g(x)<\epsilon\pi$, $\epsilon$ is the amplitude of randomness, it is to be noted that even optimally spaced binary distribution is enough for observing such phenomenon (see supplementary Section 2)] for a simple understanding of the phenomenon. Hence, the problem is reduced to one-dimension (here $x$), and the spin-orbit coupled modes are only seen along the $k_x$ axis. The momentum scale is normalized with wavevector ($k_o$) of incident Gaussian beam here and, for all further discussions. A completely symmetric ($I_t^{LCP}(k_x )=I_t^{RCP}(-k_x ))$, $I_{sa}=0$) spin-orbit coupled random spin-split scattering modes in the momentum space is obtained for the LCP and RCP polarization state. 

The real and imaginary parts of the two-point spatial autocorrelation of the transmitted beam are shown as a function of distance between the points ($\Delta x$) in Fig.\ref{fig1}d. The autocorrelation depends on the randomness of the spatial phase distribution and the width of input beam. Specifically, the real and the imaginary components have similar magnitude for random spin split modes, but for the usual spin-Hall effect  the real part gets a Gaussian nature (as that of the incident beam) with larger magnitude than the imaginary part (See supplementary section 1). The difference of the maximum of the real and the imaginary parts of the autocorrelation function with a varying width ($w_o$) of the input beam and the amplitude of randomness ($\epsilon$) were recorded for over $100$ samples. The most frequent amplitude difference for the given beam width and the amplitude of randomness is shown in Fig.\ref{fig2}. Below dotted line, the random modes of sufficiently large amplitudes (at least $0.15$ times the maximum intensity) are observed in momentum space intensity distribution in more than half cases of the system described by the parameters: the amplitude of randomness and the beam width.
\begin{figure*}[ht]
	\centering
	\includegraphics[width=0.8\linewidth]{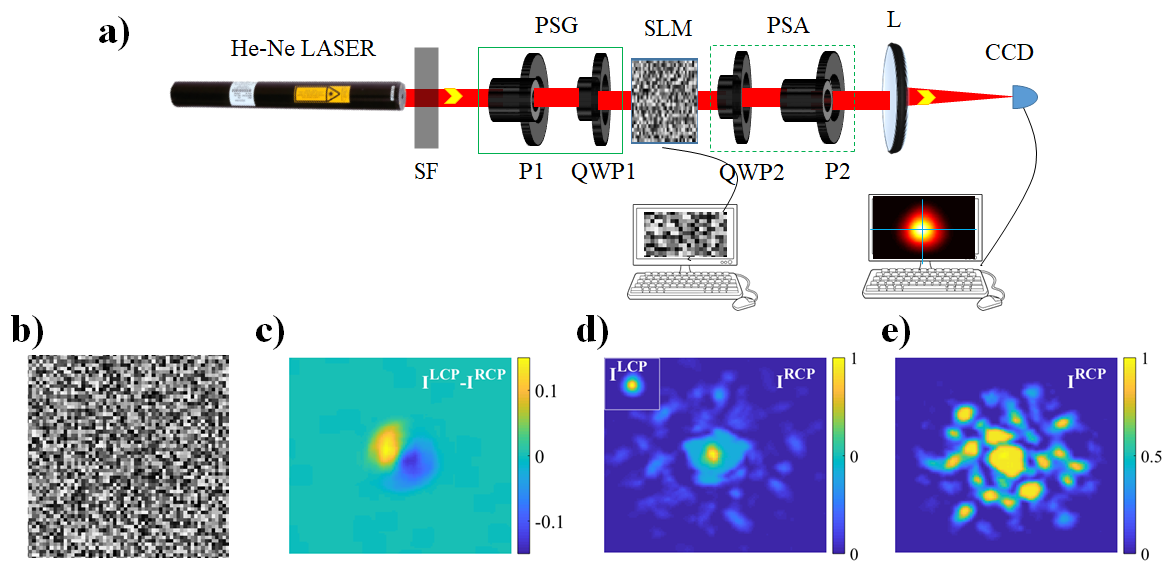}
	\caption{\label{fig4}(a) The experimental setup: a spatial filtered (SF) Gaussian beam (Helium-Neon) was passed through a polarization state generator (PSG) consisting of a polarizer (P1) and a quarter-wave plate (QWP1) to generate a linearly polarized light. The linearly polarized light was incident on a transmissive SLM whose phase distribution was controlled using a gray level image (as shown in (b)). The  transmitted light was projected on LCP and RCP polarization states using a polarization state analyzer (PSA) unit. The momentum distribution of LCP and RCP states was measured using a photodetector (CCD) kept at focal plane of the lens (L). (c) The difference between the intensity distribution for LCP and RCP polarization projection for $\epsilon_{slm}=0.1$, the usual momentum domain spin - Hall effect of light, for $8\times 8$ bins.   The momentum space intensity distribution for the RCP polarization projection showing  the random scattering modes for (d) $8\times 8$ bins (the inset shows no effect of the disorder on the LCP polarization projection) and (e) $15\times 15$ bins, in the SLM.}
\end{figure*}

The symmetricity of spin-split modes in Fourier space is guaranteed if the total phase distribution has only the geometrical phase contribution. An additional spatially disordered dynamical phase breaks symmetricity of the spin-orbit coupled random spin-split scattering modes and, an \textit{asymmetric} scattering modes from the disordered anisotropic medium ($I_{sa}>0$) is observed. The \textit{asymmetric} scattering from a disordered medium can be used to actively tune the random split modes by using an elliptical polarization state of light to deferentially excite the LCP and RCP scattering modes, as shown in Fig.\ref{fig3}a-d. However, when the geometrical and dynamical phase acquired from the disordered system is synchronous and of similar magnitude (e.g. $\phi_d(x)\approx \phi_g (x)+const.$), then the randomly scattered modes are observed for RCP while LCP shows no effect of the disordered medium. This \textit{extraordinary asymmetric} random scattering modes (Fig.\ref{fig3}a,c) arise due to a zero total phase ($\pm\phi_g (x)+\phi_d$) gradient for LCP  polarization state while the RCP gets a phase gradient twice of the acquired geometric phase gradient, as can be noted from frequency distribution of the total phase value for RCP and LCP polarization states (\ref{fig3}e,g,i,k). An asynchronous dynamical and geometrical phase distribution gives the \textit{asymmetric} random scattered momentum modes for all the elliptical polarization states, as shown in Fig.\ref{fig3}b,d.

We used the above ideas to tune the width of input beam and  the phase distribution to observe the spin selective/ asymmetric random scattering modes in an optical system. This can indeed be done in a twisted nematic liquid crystal based transmissive spatial light modulator (SLM, Holoeye LC 2002). The effective retardance value and the orientation angle (the Panchratanam Berry geometric phase) of each equivalent retarders can be simultaneously tuned in the SLM by varying the voltage applied (projected gray level value ($n$) on the SLM) across the pixels \cite{pal2016tunable} in a synchronous manner. The maximum amplitude of randomness that can be reached in such a system is quite less, as the phase retardance of SLM can be varied from $\pi/7$ to $\approx 3\pi/4$ rad and the effective orientation can be simultaneously tuned from
$0$ to $2\pi/5$ rad (see supplementary section 3) \cite{pal2016tunable}. However, the randomly scattered modes from the SLM can be observed with a beam of sufficiently small effective beam width. Thus, the $8\times 8$ pixel bins (pixel pitch $\approx 32 \mu m$) were assigned with a single value of the gray level to reduce the effective beam width ($w_o/$(\textit{size of the bin})). The gray level values ($n=60*f(\epsilon_{slm})+80$, $f(\epsilon_{slm})$ is a uniformly random distribution between $-\epsilon_{slm}$ to $\epsilon_{slm}$) were given on the SLM's pixel bins to generate a synchronous disordered anisotropic medium with random retardance value and orientation angles of the equivalent retarder.

A linearly polarized Gaussian beam (wavelength $\lambda \approx 633$ nm, $w_o=2.25$ mm or $9$ effective units w.r.t. the $8\times 8$ bin size) was incident on the SLM with a random phase distribution (random local phase gradient across each bin, a sample image with $\epsilon_{slm}=1$ is shown in Fig.\ref{fig4}b), and the transmitted light was analyzed using LCP and RCP polarization projections. The  projected beam was focused using a lens to observe the momentum space distribution at the focal plane. A small disordered in gray level distribution on SLM was given to observe the spin-Hall effect of light (Fig.\ref{fig4}c). An increase in the amplitude of randomness gives rise to the random scattering modes in the momentum space distribution for RCP polarization, as shown in Fig.\ref{fig4}d for $\epsilon_{slm}=1$. The LCP projection give no such split modes for the same distribution (inset Fig.\ref{fig4}d). It shows the extraordinary spin selectivity/asymmetricty of the scattered random modes observed due to  nearly same magnitude and synchronous nature of the dynamical and geometrical phase acquired in the SLM.  As expected, the reduced effective beam width gives rise to more dominant splited modes in momentum space (Fig.\ref{fig4}e).

It is important to note that even for maximum possible spatial randomness in the distribution of gray level of the SLM, the range of geometric phase distribution is small because of the limited range the retardance value and the orientation angle of the spatially tunable retarders in the SLM (hence the reduced effective beam width of the incident beam, see supplementary section 1).  Thus the number of split modes observed in the momentum space is comparatively small. However, such effects can be enhanced by a suitable fabrication of the liquid crystal device \cite{duran2005cell,lu1990theory}. Additionally, tailoring the incident beam polarization or spatial profile can be used to modulate such effects \cite{bender2018customizing,ling2014realization}. 

Never-the-less, it is clear from Fig.\ref{fig4} that the spin selective/asymmetric random scattering modes can be observed from a disordered anisotropic medium with synchronous dynamical and geometrical phase distributions. The finding can open up various spin-controlled application in electronics, quantum and condensed matter systems. In particular, the controlled active tuning of the randomly scattered modes using polarization of light may have potential applications in localization of ultracold atoms and trapping microparticles in a disordered potential \cite{jendrzejewski2012three,volpe2014speckle}, super resolution imaging techniques \cite{chaigne2016super}, and dynamic speckle illumination microscopy\cite{gateau2013improving}.
\bibliographystyle{apsrev4-1}
\bibliography{radialref.bib}

\begin{thebibliography}{26}%
\makeatletter
\providecommand \@ifxundefined [1]{%
 \@ifx{#1\undefined}
}%
\providecommand \@ifnum [1]{%
 \ifnum #1\expandafter \@firstoftwo
 \else \expandafter \@secondoftwo
 \fi
}%
\providecommand \@ifx [1]{%
 \ifx #1\expandafter \@firstoftwo
 \else \expandafter \@secondoftwo
 \fi
}%
\providecommand \natexlab [1]{#1}%
\providecommand \enquote  [1]{``#1''}%
\providecommand \bibnamefont  [1]{#1}%
\providecommand \bibfnamefont [1]{#1}%
\providecommand \citenamefont [1]{#1}%
\providecommand \href@noop [0]{\@secondoftwo}%
\providecommand \href [0]{\begingroup \@sanitize@url \@href}%
\providecommand \@href[1]{\@@startlink{#1}\@@href}%
\providecommand \@@href[1]{\endgroup#1\@@endlink}%
\providecommand \@sanitize@url [0]{\catcode `\\12\catcode `\$12\catcode
  `\&12\catcode `\#12\catcode `\^12\catcode `\_12\catcode `\%12\relax}%
\providecommand \@@startlink[1]{}%
\providecommand \@@endlink[0]{}%
\providecommand \url  [0]{\begingroup\@sanitize@url \@url }%
\providecommand \@url [1]{\endgroup\@href {#1}{\urlprefix }}%
\providecommand \urlprefix  [0]{URL }%
\providecommand \Eprint [0]{\href }%
\providecommand \doibase [0]{http://dx.doi.org/}%
\providecommand \selectlanguage [0]{\@gobble}%
\providecommand \bibinfo  [0]{\@secondoftwo}%
\providecommand \bibfield  [0]{\@secondoftwo}%
\providecommand \translation [1]{[#1]}%
\providecommand \BibitemOpen [0]{}%
\providecommand \bibitemStop [0]{}%
\providecommand \bibitemNoStop [0]{.\EOS\space}%
\providecommand \EOS [0]{\spacefactor3000\relax}%
\providecommand \BibitemShut  [1]{\csname bibitem#1\endcsname}%
\let\auto@bib@innerbib\@empty
\bibitem [{\citenamefont {Bliokh}\ \emph {et~al.}(2015)\citenamefont {Bliokh},
  \citenamefont {Rodr{\'\i}guez-Fortu{\~n}o}, \citenamefont {Nori},\ and\
  \citenamefont {Zayats}}]{bliokh2015spin}%
  \BibitemOpen
  \bibfield  {author} {\bibinfo {author} {\bibfnamefont {K.~Y.}\ \bibnamefont
  {Bliokh}}, \bibinfo {author} {\bibfnamefont {F.~J.}\ \bibnamefont
  {Rodr{\'\i}guez-Fortu{\~n}o}}, \bibinfo {author} {\bibfnamefont
  {F.}~\bibnamefont {Nori}}, \ and\ \bibinfo {author} {\bibfnamefont {A.~V.}\
  \bibnamefont {Zayats}},\ }\href@noop {} {\bibfield  {journal} {\bibinfo
  {journal} {Nature Photonics}\ }\textbf {\bibinfo {volume} {9}},\ \bibinfo
  {pages} {796} (\bibinfo {year} {2015})}\BibitemShut {NoStop}%
\bibitem [{\citenamefont {Ma}\ \emph {et~al.}(2016)\citenamefont {Ma},
  \citenamefont {Li}, \citenamefont {Fomin}, \citenamefont {Hentschel},
  \citenamefont {G{\"o}tte}, \citenamefont {Yin}, \citenamefont {Jorgensen},\
  and\ \citenamefont {Schmidt}}]{ma2016spin}%
  \BibitemOpen
  \bibfield  {author} {\bibinfo {author} {\bibfnamefont {L.}~\bibnamefont
  {Ma}}, \bibinfo {author} {\bibfnamefont {S.}~\bibnamefont {Li}}, \bibinfo
  {author} {\bibfnamefont {V.~M.}\ \bibnamefont {Fomin}}, \bibinfo {author}
  {\bibfnamefont {M.}~\bibnamefont {Hentschel}}, \bibinfo {author}
  {\bibfnamefont {J.~B.}\ \bibnamefont {G{\"o}tte}}, \bibinfo {author}
  {\bibfnamefont {Y.}~\bibnamefont {Yin}}, \bibinfo {author} {\bibfnamefont
  {M.}~\bibnamefont {Jorgensen}}, \ and\ \bibinfo {author} {\bibfnamefont
  {O.~G.}\ \bibnamefont {Schmidt}},\ }\href@noop {} {\bibfield  {journal}
  {\bibinfo  {journal} {Nature communications}\ }\textbf {\bibinfo {volume}
  {7}},\ \bibinfo {pages} {10983} (\bibinfo {year} {2016})}\BibitemShut
  {NoStop}%
\bibitem [{\citenamefont {Shitrit}\ \emph
  {et~al.}(2013{\natexlab{a}})\citenamefont {Shitrit}, \citenamefont {Maayani},
  \citenamefont {Veksler}, \citenamefont {Kleiner},\ and\ \citenamefont
  {Hasman}}]{shitrit2013rashba}%
  \BibitemOpen
  \bibfield  {author} {\bibinfo {author} {\bibfnamefont {N.}~\bibnamefont
  {Shitrit}}, \bibinfo {author} {\bibfnamefont {S.}~\bibnamefont {Maayani}},
  \bibinfo {author} {\bibfnamefont {D.}~\bibnamefont {Veksler}}, \bibinfo
  {author} {\bibfnamefont {V.}~\bibnamefont {Kleiner}}, \ and\ \bibinfo
  {author} {\bibfnamefont {E.}~\bibnamefont {Hasman}},\ }\href@noop {}
  {\bibfield  {journal} {\bibinfo  {journal} {Optics letters}\ }\textbf
  {\bibinfo {volume} {38}},\ \bibinfo {pages} {4358} (\bibinfo {year}
  {2013}{\natexlab{a}})}\BibitemShut {NoStop}%
\bibitem [{\citenamefont {Shitrit}\ \emph
  {et~al.}(2013{\natexlab{b}})\citenamefont {Shitrit}, \citenamefont
  {Yulevich}, \citenamefont {Maguid}, \citenamefont {Ozeri}, \citenamefont
  {Veksler}, \citenamefont {Kleiner},\ and\ \citenamefont
  {Hasman}}]{shitrit2013spin}%
  \BibitemOpen
  \bibfield  {author} {\bibinfo {author} {\bibfnamefont {N.}~\bibnamefont
  {Shitrit}}, \bibinfo {author} {\bibfnamefont {I.}~\bibnamefont {Yulevich}},
  \bibinfo {author} {\bibfnamefont {E.}~\bibnamefont {Maguid}}, \bibinfo
  {author} {\bibfnamefont {D.}~\bibnamefont {Ozeri}}, \bibinfo {author}
  {\bibfnamefont {D.}~\bibnamefont {Veksler}}, \bibinfo {author} {\bibfnamefont
  {V.}~\bibnamefont {Kleiner}}, \ and\ \bibinfo {author} {\bibfnamefont
  {E.}~\bibnamefont {Hasman}},\ }\href@noop {} {\bibfield  {journal} {\bibinfo
  {journal} {Science}\ }\textbf {\bibinfo {volume} {340}},\ \bibinfo {pages}
  {724} (\bibinfo {year} {2013}{\natexlab{b}})}\BibitemShut {NoStop}%
\bibitem [{\citenamefont {Neugebauer}\ \emph {et~al.}(2014)\citenamefont
  {Neugebauer}, \citenamefont {Banzer}, \citenamefont {Bauer}, \citenamefont
  {Orlov}, \citenamefont {Lindlein}, \citenamefont {Aiello},\ and\
  \citenamefont {Leuchs}}]{gerd_2014}%
  \BibitemOpen
  \bibfield  {author} {\bibinfo {author} {\bibfnamefont {M.}~\bibnamefont
  {Neugebauer}}, \bibinfo {author} {\bibfnamefont {P.}~\bibnamefont {Banzer}},
  \bibinfo {author} {\bibfnamefont {T.}~\bibnamefont {Bauer}}, \bibinfo
  {author} {\bibfnamefont {S.}~\bibnamefont {Orlov}}, \bibinfo {author}
  {\bibfnamefont {N.}~\bibnamefont {Lindlein}}, \bibinfo {author}
  {\bibfnamefont {A.}~\bibnamefont {Aiello}}, \ and\ \bibinfo {author}
  {\bibfnamefont {G.}~\bibnamefont {Leuchs}},\ }\href {\doibase
  10.1103/physreva.89.013840} {\bibfield  {journal} {\bibinfo  {journal} {Phys.
  Rev. A}\ }\textbf {\bibinfo {volume} {89}} (\bibinfo {year} {2014}),\
  10.1103/physreva.89.013840}\BibitemShut {NoStop}%
\bibitem [{\citenamefont {Lin}\ \emph {et~al.}(2014)\citenamefont {Lin},
  \citenamefont {Fan}, \citenamefont {Hasman},\ and\ \citenamefont
  {Brongersma}}]{lin2014dielectric}%
  \BibitemOpen
  \bibfield  {author} {\bibinfo {author} {\bibfnamefont {D.}~\bibnamefont
  {Lin}}, \bibinfo {author} {\bibfnamefont {P.}~\bibnamefont {Fan}}, \bibinfo
  {author} {\bibfnamefont {E.}~\bibnamefont {Hasman}}, \ and\ \bibinfo {author}
  {\bibfnamefont {M.~L.}\ \bibnamefont {Brongersma}},\ }\href@noop {}
  {\bibfield  {journal} {\bibinfo  {journal} {science}\ }\textbf {\bibinfo
  {volume} {345}},\ \bibinfo {pages} {298} (\bibinfo {year}
  {2014})}\BibitemShut {NoStop}%
\bibitem [{\citenamefont {Ling}\ \emph {et~al.}(2014)\citenamefont {Ling},
  \citenamefont {Zhou}, \citenamefont {Shu}, \citenamefont {Luo},\ and\
  \citenamefont {Wen}}]{ling2014realization}%
  \BibitemOpen
  \bibfield  {author} {\bibinfo {author} {\bibfnamefont {X.}~\bibnamefont
  {Ling}}, \bibinfo {author} {\bibfnamefont {X.}~\bibnamefont {Zhou}}, \bibinfo
  {author} {\bibfnamefont {W.}~\bibnamefont {Shu}}, \bibinfo {author}
  {\bibfnamefont {H.}~\bibnamefont {Luo}}, \ and\ \bibinfo {author}
  {\bibfnamefont {S.}~\bibnamefont {Wen}},\ }\href@noop {} {\bibfield
  {journal} {\bibinfo  {journal} {Scientific reports}\ }\textbf {\bibinfo
  {volume} {4}},\ \bibinfo {pages} {5557} (\bibinfo {year} {2014})}\BibitemShut
  {NoStop}%
\bibitem [{\citenamefont {Ling}\ \emph {et~al.}(2015)\citenamefont {Ling},
  \citenamefont {Zhou}, \citenamefont {Yi}, \citenamefont {Shu}, \citenamefont
  {Liu}, \citenamefont {Chen}, \citenamefont {Luo}, \citenamefont {Wen},\ and\
  \citenamefont {Fan}}]{ling2015giant}%
  \BibitemOpen
  \bibfield  {author} {\bibinfo {author} {\bibfnamefont {X.}~\bibnamefont
  {Ling}}, \bibinfo {author} {\bibfnamefont {X.}~\bibnamefont {Zhou}}, \bibinfo
  {author} {\bibfnamefont {X.}~\bibnamefont {Yi}}, \bibinfo {author}
  {\bibfnamefont {W.}~\bibnamefont {Shu}}, \bibinfo {author} {\bibfnamefont
  {Y.}~\bibnamefont {Liu}}, \bibinfo {author} {\bibfnamefont {S.}~\bibnamefont
  {Chen}}, \bibinfo {author} {\bibfnamefont {H.}~\bibnamefont {Luo}}, \bibinfo
  {author} {\bibfnamefont {S.}~\bibnamefont {Wen}}, \ and\ \bibinfo {author}
  {\bibfnamefont {D.}~\bibnamefont {Fan}},\ }\href@noop {} {\bibfield
  {journal} {\bibinfo  {journal} {Light: Science \& Applications}\ }\textbf
  {\bibinfo {volume} {4}},\ \bibinfo {pages} {e290} (\bibinfo {year}
  {2015})}\BibitemShut {NoStop}%
\bibitem [{\citenamefont {Pal}\ \emph {et~al.}(2016)\citenamefont {Pal},
  \citenamefont {Banerjee}, \citenamefont {Chandel}, \citenamefont {Bag},
  \citenamefont {Majumder},\ and\ \citenamefont {Ghosh}}]{pal2016tunable}%
  \BibitemOpen
  \bibfield  {author} {\bibinfo {author} {\bibfnamefont {M.}~\bibnamefont
  {Pal}}, \bibinfo {author} {\bibfnamefont {C.}~\bibnamefont {Banerjee}},
  \bibinfo {author} {\bibfnamefont {S.}~\bibnamefont {Chandel}}, \bibinfo
  {author} {\bibfnamefont {A.}~\bibnamefont {Bag}}, \bibinfo {author}
  {\bibfnamefont {S.~K.}\ \bibnamefont {Majumder}}, \ and\ \bibinfo {author}
  {\bibfnamefont {N.}~\bibnamefont {Ghosh}},\ }\href@noop {} {\bibfield
  {journal} {\bibinfo  {journal} {Scientific reports}\ }\textbf {\bibinfo
  {volume} {6}},\ \bibinfo {pages} {39582} (\bibinfo {year}
  {2016})}\BibitemShut {NoStop}%
\bibitem [{\citenamefont {Ling}\ \emph {et~al.}(2017)\citenamefont {Ling},
  \citenamefont {Zhou}, \citenamefont {Huang}, \citenamefont {Liu},
  \citenamefont {Qiu}, \citenamefont {Luo},\ and\ \citenamefont
  {Wen}}]{ling2017recent}%
  \BibitemOpen
  \bibfield  {author} {\bibinfo {author} {\bibfnamefont {X.}~\bibnamefont
  {Ling}}, \bibinfo {author} {\bibfnamefont {X.}~\bibnamefont {Zhou}}, \bibinfo
  {author} {\bibfnamefont {K.}~\bibnamefont {Huang}}, \bibinfo {author}
  {\bibfnamefont {Y.}~\bibnamefont {Liu}}, \bibinfo {author} {\bibfnamefont
  {C.-W.}\ \bibnamefont {Qiu}}, \bibinfo {author} {\bibfnamefont
  {H.}~\bibnamefont {Luo}}, \ and\ \bibinfo {author} {\bibfnamefont
  {S.}~\bibnamefont {Wen}},\ }\href@noop {} {\bibfield  {journal} {\bibinfo
  {journal} {Reports on Progress in Physics}\ }\textbf {\bibinfo {volume}
  {80}},\ \bibinfo {pages} {066401} (\bibinfo {year} {2017})}\BibitemShut
  {NoStop}%
\bibitem [{\citenamefont {Hosten}\ and\ \citenamefont
  {Kwiat}(2008)}]{hosten2008observation}%
  \BibitemOpen
  \bibfield  {author} {\bibinfo {author} {\bibfnamefont {O.}~\bibnamefont
  {Hosten}}\ and\ \bibinfo {author} {\bibfnamefont {P.}~\bibnamefont {Kwiat}},\
  }\href@noop {} {\bibfield  {journal} {\bibinfo  {journal} {Science}\ }\textbf
  {\bibinfo {volume} {319}},\ \bibinfo {pages} {787} (\bibinfo {year}
  {2008})}\BibitemShut {NoStop}%
\bibitem [{\citenamefont {Maguid}\ \emph {et~al.}(2017)\citenamefont {Maguid},
  \citenamefont {Yannai}, \citenamefont {Faerman}, \citenamefont {Yulevich},
  \citenamefont {Kleiner},\ and\ \citenamefont {Hasman}}]{maguid2017disorder}%
  \BibitemOpen
  \bibfield  {author} {\bibinfo {author} {\bibfnamefont {E.}~\bibnamefont
  {Maguid}}, \bibinfo {author} {\bibfnamefont {M.}~\bibnamefont {Yannai}},
  \bibinfo {author} {\bibfnamefont {A.}~\bibnamefont {Faerman}}, \bibinfo
  {author} {\bibfnamefont {I.}~\bibnamefont {Yulevich}}, \bibinfo {author}
  {\bibfnamefont {V.}~\bibnamefont {Kleiner}}, \ and\ \bibinfo {author}
  {\bibfnamefont {E.}~\bibnamefont {Hasman}},\ }\href@noop {} {\bibfield
  {journal} {\bibinfo  {journal} {Science}\ }\textbf {\bibinfo {volume}
  {358}},\ \bibinfo {pages} {1411} (\bibinfo {year} {2017})}\BibitemShut
  {NoStop}%
\bibitem [{\citenamefont {Frischwasser}\ \emph {et~al.}(2011)\citenamefont
  {Frischwasser}, \citenamefont {Yulevich}, \citenamefont {Kleiner},\ and\
  \citenamefont {Hasman}}]{frischwasser2011rashba}%
  \BibitemOpen
  \bibfield  {author} {\bibinfo {author} {\bibfnamefont {K.}~\bibnamefont
  {Frischwasser}}, \bibinfo {author} {\bibfnamefont {I.}~\bibnamefont
  {Yulevich}}, \bibinfo {author} {\bibfnamefont {V.}~\bibnamefont {Kleiner}}, \
  and\ \bibinfo {author} {\bibfnamefont {E.}~\bibnamefont {Hasman}},\
  }\href@noop {} {\bibfield  {journal} {\bibinfo  {journal} {Optics express}\
  }\textbf {\bibinfo {volume} {19}},\ \bibinfo {pages} {23475} (\bibinfo {year}
  {2011})}\BibitemShut {NoStop}%
\bibitem [{\citenamefont {Bliokh}\ \emph {et~al.}(2007)\citenamefont {Bliokh},
  \citenamefont {Frolov},\ and\ \citenamefont {Kravtsov}}]{PhysRevA.75.053821}%
  \BibitemOpen
  \bibfield  {author} {\bibinfo {author} {\bibfnamefont {K.~Y.}\ \bibnamefont
  {Bliokh}}, \bibinfo {author} {\bibfnamefont {D.~Y.}\ \bibnamefont {Frolov}},
  \ and\ \bibinfo {author} {\bibfnamefont {Y.~A.}\ \bibnamefont {Kravtsov}},\
  }\href {\doibase 10.1103/PhysRevA.75.053821} {\bibfield  {journal} {\bibinfo
  {journal} {Phys. Rev. A}\ }\textbf {\bibinfo {volume} {75}},\ \bibinfo
  {pages} {053821} (\bibinfo {year} {2007})}\BibitemShut {NoStop}%
\bibitem [{\citenamefont {Wilczek}\ and\ \citenamefont
  {Zee}(1984)}]{wilczek1984appearance}%
  \BibitemOpen
  \bibfield  {author} {\bibinfo {author} {\bibfnamefont {F.}~\bibnamefont
  {Wilczek}}\ and\ \bibinfo {author} {\bibfnamefont {A.}~\bibnamefont {Zee}},\
  }\href@noop {} {\bibfield  {journal} {\bibinfo  {journal} {Physical Review
  Letters}\ }\textbf {\bibinfo {volume} {52}},\ \bibinfo {pages} {2111}
  (\bibinfo {year} {1984})}\BibitemShut {NoStop}%
\bibitem [{\citenamefont {Zak}(1989)}]{zak1989berry}%
  \BibitemOpen
  \bibfield  {author} {\bibinfo {author} {\bibfnamefont {J.}~\bibnamefont
  {Zak}},\ }\href@noop {} {\bibfield  {journal} {\bibinfo  {journal} {EPL
  (Europhysics Letters)}\ }\textbf {\bibinfo {volume} {9}},\ \bibinfo {pages}
  {615} (\bibinfo {year} {1989})}\BibitemShut {NoStop}%
\bibitem [{\citenamefont {de~Polavieja}(1998)}]{de1998noncyclic}%
  \BibitemOpen
  \bibfield  {author} {\bibinfo {author} {\bibfnamefont {G.~G.}\ \bibnamefont
  {de~Polavieja}},\ }\href@noop {} {\bibfield  {journal} {\bibinfo  {journal}
  {Physical review letters}\ }\textbf {\bibinfo {volume} {81}},\ \bibinfo
  {pages} {1} (\bibinfo {year} {1998})}\BibitemShut {NoStop}%
\bibitem [{\citenamefont {Manchon}\ \emph {et~al.}(2015)\citenamefont
  {Manchon}, \citenamefont {Koo}, \citenamefont {Nitta}, \citenamefont
  {Frolov},\ and\ \citenamefont {Duine}}]{manchon2015new}%
  \BibitemOpen
  \bibfield  {author} {\bibinfo {author} {\bibfnamefont {A.}~\bibnamefont
  {Manchon}}, \bibinfo {author} {\bibfnamefont {H.~C.}\ \bibnamefont {Koo}},
  \bibinfo {author} {\bibfnamefont {J.}~\bibnamefont {Nitta}}, \bibinfo
  {author} {\bibfnamefont {S.}~\bibnamefont {Frolov}}, \ and\ \bibinfo {author}
  {\bibfnamefont {R.}~\bibnamefont {Duine}},\ }\href@noop {} {\bibfield
  {journal} {\bibinfo  {journal} {Nature materials}\ }\textbf {\bibinfo
  {volume} {14}},\ \bibinfo {pages} {871} (\bibinfo {year} {2015})}\BibitemShut
  {NoStop}%
\bibitem [{\citenamefont {Jendrzejewski}\ \emph {et~al.}(2012)\citenamefont
  {Jendrzejewski}, \citenamefont {Bernard}, \citenamefont {Mueller},
  \citenamefont {Cheinet}, \citenamefont {Josse}, \citenamefont {Piraud},
  \citenamefont {Pezz{\'e}}, \citenamefont {Sanchez-Palencia}, \citenamefont
  {Aspect},\ and\ \citenamefont {Bouyer}}]{jendrzejewski2012three}%
  \BibitemOpen
  \bibfield  {author} {\bibinfo {author} {\bibfnamefont {F.}~\bibnamefont
  {Jendrzejewski}}, \bibinfo {author} {\bibfnamefont {A.}~\bibnamefont
  {Bernard}}, \bibinfo {author} {\bibfnamefont {K.}~\bibnamefont {Mueller}},
  \bibinfo {author} {\bibfnamefont {P.}~\bibnamefont {Cheinet}}, \bibinfo
  {author} {\bibfnamefont {V.}~\bibnamefont {Josse}}, \bibinfo {author}
  {\bibfnamefont {M.}~\bibnamefont {Piraud}}, \bibinfo {author} {\bibfnamefont
  {L.}~\bibnamefont {Pezz{\'e}}}, \bibinfo {author} {\bibfnamefont
  {L.}~\bibnamefont {Sanchez-Palencia}}, \bibinfo {author} {\bibfnamefont
  {A.}~\bibnamefont {Aspect}}, \ and\ \bibinfo {author} {\bibfnamefont
  {P.}~\bibnamefont {Bouyer}},\ }\href@noop {} {\bibfield  {journal} {\bibinfo
  {journal} {Nature Physics}\ }\textbf {\bibinfo {volume} {8}},\ \bibinfo
  {pages} {398} (\bibinfo {year} {2012})}\BibitemShut {NoStop}%
\bibitem [{\citenamefont {Volpe}\ \emph {et~al.}(2014)\citenamefont {Volpe},
  \citenamefont {Kurz}, \citenamefont {Callegari}, \citenamefont {Volpe},\ and\
  \citenamefont {Gigan}}]{volpe2014speckle}%
  \BibitemOpen
  \bibfield  {author} {\bibinfo {author} {\bibfnamefont {G.}~\bibnamefont
  {Volpe}}, \bibinfo {author} {\bibfnamefont {L.}~\bibnamefont {Kurz}},
  \bibinfo {author} {\bibfnamefont {A.}~\bibnamefont {Callegari}}, \bibinfo
  {author} {\bibfnamefont {G.}~\bibnamefont {Volpe}}, \ and\ \bibinfo {author}
  {\bibfnamefont {S.}~\bibnamefont {Gigan}},\ }\href@noop {} {\bibfield
  {journal} {\bibinfo  {journal} {Optics express}\ }\textbf {\bibinfo {volume}
  {22}},\ \bibinfo {pages} {18159} (\bibinfo {year} {2014})}\BibitemShut
  {NoStop}%
\bibitem [{\citenamefont {Chaigne}\ \emph {et~al.}(2016)\citenamefont
  {Chaigne}, \citenamefont {Gateau}, \citenamefont {Allain}, \citenamefont
  {Katz}, \citenamefont {Gigan}, \citenamefont {Sentenac},\ and\ \citenamefont
  {Bossy}}]{chaigne2016super}%
  \BibitemOpen
  \bibfield  {author} {\bibinfo {author} {\bibfnamefont {T.}~\bibnamefont
  {Chaigne}}, \bibinfo {author} {\bibfnamefont {J.}~\bibnamefont {Gateau}},
  \bibinfo {author} {\bibfnamefont {M.}~\bibnamefont {Allain}}, \bibinfo
  {author} {\bibfnamefont {O.}~\bibnamefont {Katz}}, \bibinfo {author}
  {\bibfnamefont {S.}~\bibnamefont {Gigan}}, \bibinfo {author} {\bibfnamefont
  {A.}~\bibnamefont {Sentenac}}, \ and\ \bibinfo {author} {\bibfnamefont
  {E.}~\bibnamefont {Bossy}},\ }\href@noop {} {\bibfield  {journal} {\bibinfo
  {journal} {Optica}\ }\textbf {\bibinfo {volume} {3}},\ \bibinfo {pages} {54}
  (\bibinfo {year} {2016})}\BibitemShut {NoStop}%
\bibitem [{\citenamefont {Gateau}\ \emph {et~al.}(2013)\citenamefont {Gateau},
  \citenamefont {Chaigne}, \citenamefont {Katz}, \citenamefont {Gigan},\ and\
  \citenamefont {Bossy}}]{gateau2013improving}%
  \BibitemOpen
  \bibfield  {author} {\bibinfo {author} {\bibfnamefont {J.}~\bibnamefont
  {Gateau}}, \bibinfo {author} {\bibfnamefont {T.}~\bibnamefont {Chaigne}},
  \bibinfo {author} {\bibfnamefont {O.}~\bibnamefont {Katz}}, \bibinfo {author}
  {\bibfnamefont {S.}~\bibnamefont {Gigan}}, \ and\ \bibinfo {author}
  {\bibfnamefont {E.}~\bibnamefont {Bossy}},\ }\href@noop {} {\bibfield
  {journal} {\bibinfo  {journal} {Optics letters}\ }\textbf {\bibinfo {volume}
  {38}},\ \bibinfo {pages} {5188} (\bibinfo {year} {2013})}\BibitemShut
  {NoStop}%
\bibitem [{\citenamefont {Gupta}\ \emph {et~al.}(2015)\citenamefont {Gupta},
  \citenamefont {Ghosh},\ and\ \citenamefont {Banerjee}}]{ngbook}%
  \BibitemOpen
  \bibfield  {author} {\bibinfo {author} {\bibfnamefont {S.~D.}\ \bibnamefont
  {Gupta}}, \bibinfo {author} {\bibfnamefont {N.}~\bibnamefont {Ghosh}}, \ and\
  \bibinfo {author} {\bibfnamefont {A.}~\bibnamefont {Banerjee}},\ }\href@noop
  {} {\emph {\bibinfo {title} {Wave Optics: Basic Concepts and Contemporary
  Trends}}}\ (\bibinfo  {publisher} {CRC Press},\ \bibinfo {year}
  {2015})\BibitemShut {NoStop}%
\bibitem [{\citenamefont {Duran}\ \emph {et~al.}(2005)\citenamefont {Duran},
  \citenamefont {Lancis}, \citenamefont {Tajahuerce},\ and\ \citenamefont
  {Jaroszewicz}}]{duran2005cell}%
  \BibitemOpen
  \bibfield  {author} {\bibinfo {author} {\bibfnamefont {V.}~\bibnamefont
  {Duran}}, \bibinfo {author} {\bibfnamefont {J.}~\bibnamefont {Lancis}},
  \bibinfo {author} {\bibfnamefont {E.}~\bibnamefont {Tajahuerce}}, \ and\
  \bibinfo {author} {\bibfnamefont {Z.}~\bibnamefont {Jaroszewicz}},\
  }\href@noop {} {\bibfield  {journal} {\bibinfo  {journal} {Journal of applied
  physics}\ }\textbf {\bibinfo {volume} {97}},\ \bibinfo {pages} {043101}
  (\bibinfo {year} {2005})}\BibitemShut {NoStop}%
\bibitem [{\citenamefont {Lu}\ and\ \citenamefont
  {Saleh}(1990)}]{lu1990theory}%
  \BibitemOpen
  \bibfield  {author} {\bibinfo {author} {\bibfnamefont {K.}~\bibnamefont
  {Lu}}\ and\ \bibinfo {author} {\bibfnamefont {B.~E.}\ \bibnamefont {Saleh}},\
  }\href@noop {} {\bibfield  {journal} {\bibinfo  {journal} {Optical
  Engineering}\ }\textbf {\bibinfo {volume} {29}},\ \bibinfo {pages} {240}
  (\bibinfo {year} {1990})}\BibitemShut {NoStop}%
\bibitem [{\citenamefont {Bender}\ \emph {et~al.}(2018)\citenamefont {Bender},
  \citenamefont {Y{\i}lmaz}, \citenamefont {Bromberg},\ and\ \citenamefont
  {Cao}}]{bender2018customizing}%
  \BibitemOpen
  \bibfield  {author} {\bibinfo {author} {\bibfnamefont {N.}~\bibnamefont
  {Bender}}, \bibinfo {author} {\bibfnamefont {H.}~\bibnamefont {Y{\i}lmaz}},
  \bibinfo {author} {\bibfnamefont {Y.}~\bibnamefont {Bromberg}}, \ and\
  \bibinfo {author} {\bibfnamefont {H.}~\bibnamefont {Cao}},\ }\href@noop {}
  {\bibfield  {journal} {\bibinfo  {journal} {Optica}\ }\textbf {\bibinfo
  {volume} {5}},\ \bibinfo {pages} {595} (\bibinfo {year} {2018})}\BibitemShut
  {NoStop}%
\end{thebibliography}%

\widetext
\clearpage
\begin{center}
	\textbf{\large Supplemental Material}
\end{center}
\setcounter{equation}{0}
\setcounter{figure}{0}
\setcounter{table}{0}
\setcounter{page}{1}
\makeatletter
\renewcommand{\theequation}{S\arabic{equation}}
\renewcommand{\thefigure}{S\arabic{figure}}

\section{Section 1: The role of beam width  and autocorrelation function on momentum space intensity distribution}
\label{Section1}
\begin{figure*}[ht]
	\centering
	\includegraphics[width=0.9\linewidth]{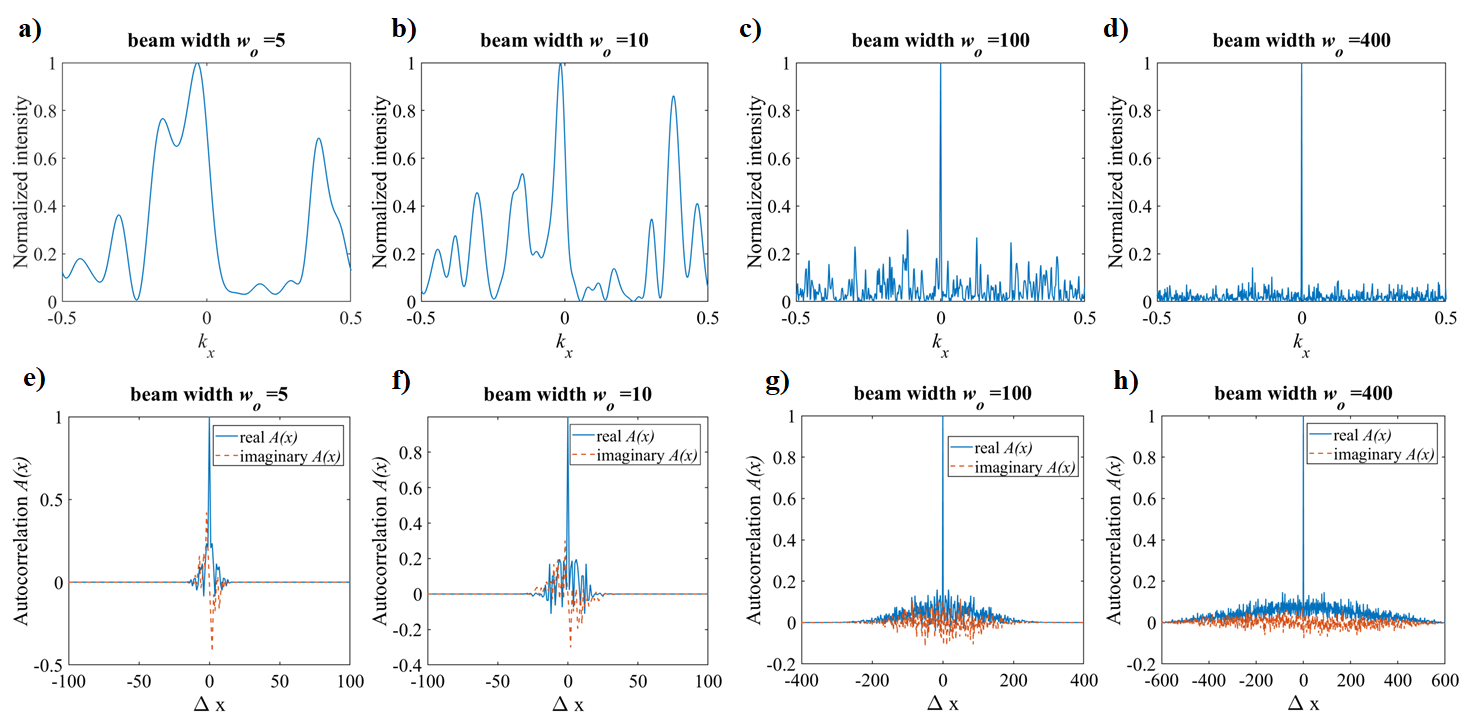}
	\caption{\label{S1}The role of beam width on momentum space (a-d) intensity distribution of the random scattering modes in momentum space (the momentum scale is normalized with wavevector ($k_o$) of the incident beam) and, (e-h) the autocorrelation function of the beam transmitted from a disordered inhomogeneous anisotropic medium, for different beam width of input Gaussian beam.}
\end{figure*}

We have shown the momentum space intensity distribution for the RCP input Gaussian beams having various values of beam width ($w_o$), as shown in Fig.\ref{S1}, for a random geometric phase distribution with $\epsilon=0.8$. It is to be noted from the figure that for same distribution of geometric phase we obtain two different limit of the effect, with the smaller beam widths showing the spin symmetric random scattering modes while the Gaussian beam with larger beam width shows the spin-Hall effect of light due to the broken inversion symmetry of the inhomogeneous anisotropic structure media. It gives clear evidence of the role of beam width of the incident light beam on the effect observed in the momentum space, i.e. the usual momentum domain spin-Hall effect of light or the spin symmetric random scattering modes. In Fig.\ref{S1}e-h, we have shown the real and imaginary part of the two-point autocorrelation function (with the distance between the points $\Delta x$) of the transmitted beam from the inhomogeneous anisotropic media for various input beam widths. A clear difference in the autocorrelation is seen for the two cases of the spin-Hall effect and the spin symmetric random scattering modes. The real and the imaginary parts show similar magnitude for the spin symmetric random scattering modes. However, a deviation between the magnitude of the real and imaginary part of the correlation is observed for the usual spin-Hall effect. Thus, the spatial autocorrelation function can be used to understand the role of the beam width of the Gaussian beam incident on the system and get a physical understanding of the critical range of the randomized phase value needed to observe the random optical Rashba effect. 

It can also be seen from the plots with the reducing beam width, the density/number of the random scattering modes observed in Fourier space reduces significantly. It leads to a small number of the random scattered modes observed from the SLM in our experiment as the effective beam width for our system is $9$ units.\\

\section{Section 2: Minimum independent elements for observing random scattering modes}
\label{Section2}
\begin{figure*}[ht]
	\centering
	\includegraphics[width=0.9\linewidth]{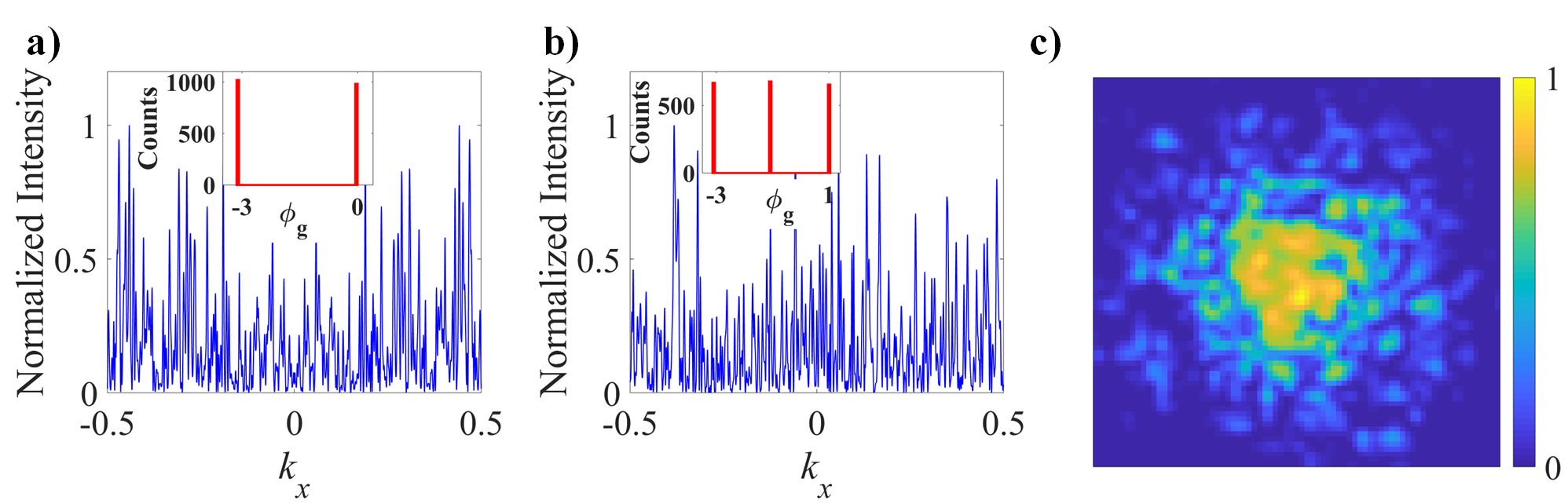}
	\caption{\label{S2}The momentum distribution of RCP polarization state along $k_x$ of the Gaussian beam transmitted from (a) only two unit cell random medium with half-wave plate oriented either at $0$ or $-\pi/2$ with lab frame throughout the random structure; (b) only three unit cell random medium with half-wave plate oriented at $0$, $-\pi/6$ or $\pi/6$ with lab frame throughout the random structure. (c) The momentum space intensity distribution for the RCP polarization projection, showing the random scattering modes for binary distribution of gray level in $8\times 8$ bins of the SLM.}
\end{figure*}

Here, we show that only two \textit{optimum} units the anisotropic elements distributed randomly throughout space are sufficient for observing random scattering modes in momentum space. Fig.\ref{S2} shows the momentum space intensity distribution of the beam transmitted from a spatially random anisotropic medium consisting of only two orientation ($-\pi/2$ and $0$, as shown in the inset Fig.\ref{S2}a) of the half-wave plate retarder distributed randomly throughout space.  In addition, we show the momentum intensity distribution obtained using only three optimum unit cells with different orientations of half-wave plate retarders. The random scattering modes were also experimentally observed in scattering form the SLM for a given random binary distribution of gray levels in for $8\times 8$ bins of the SLM pixels as shown in Fig.\ref{S2}c.

\section{Section 3: The polarization parameters of the SLM}
\label{Section3}
\begin{figure*}[ht]
	\centering
	\includegraphics[width=0.6\linewidth]{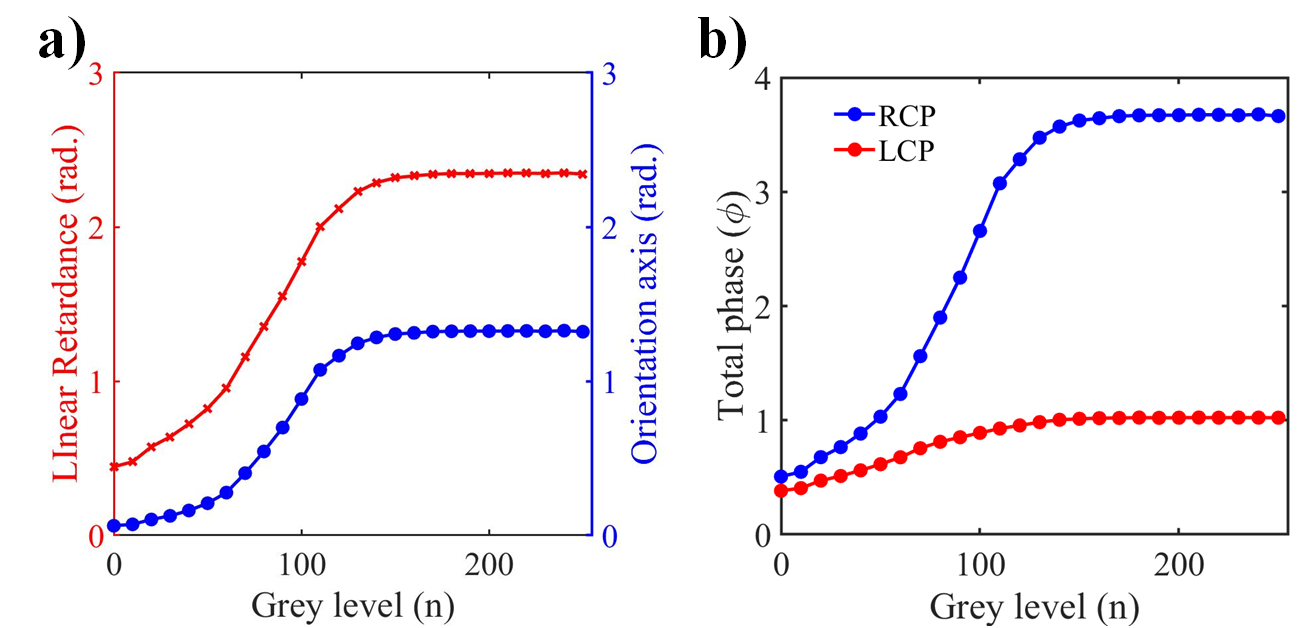}
	\caption{\label{S3}The dependence of (a) polarization parameters and (b) the total phase ($\phi_d\pm\phi_g$) occupied by the light transmitted for input LCP and RCP polarization states of light on the gray level projected on the SLM (adopted from Ref.\cite{pal2016tunable}).}
\end{figure*}
The figure \ref{S3} shows the phase retardance and the orientation of a retarder formed at a given value of the gray level in the SLM. The value of the effective retardance and the effective orientation angle of the equivalent retarder are determined by the gray value of the SLM that regulates the voltage applied across the pixels of SLM. It is seen that for a gray value in the range of $20$ to $140$ the parameters are approximately linear and, have the same slope with the SLM’s gray level. The total phase occupied for incident LCP and RCP polarization state of light with varying gray levels is shown in Fig \ref{S3}b. The total phase occupied for input RCP polarization state of light shows a significant dependence on the gray level of the SLM pixel as compared to LCP polarization state of light for which, not much deviation is found with varying gray level.

\end{document}